\documentclass[a4paper,10pt,conference]{IEEEtran}
\usepackage[frozencache=true]{minted}
\usepackage{balance}
\usepackage{graphicx}
\usepackage{color}
\usepackage[dvipsnames]{xcolor}
\usepackage{listings}
\usepackage{fancyvrb}
\usepackage{framed}
\usepackage{pgfplots}
\usepackage{fancyhdr}
\graphicspath{
    {.}
	{images/graphs/}
	{images/img_methodology/}
	{images/img_results_and_discussion/}
	{images/chapter1/}
    {figures/}
}

\ifCLASSINFOpdf
\else
\fi
\begin{document}
%
\title{An Optimized Parallel Failure-less Aho-Corasick Algorithm for DNA Sequence Matching}

\author{\IEEEauthorblockN{D.R.V.L.B Thambawita}
\IEEEauthorblockA{Department of Computer Science and Technology\\
Uva Wellassa University\\
Badulla, Sri Lanka\\
Email: vlbthambawita@gmail.com}
\and
\IEEEauthorblockN{Roshan G. Ragel and Dhammike Elkaduwe}
\IEEEauthorblockA{Department of Computer Engineering\\
University of Peradeniya\\
Peradeniya, Sri Lanka\\
Email: [ragelrg, dhammika.elkaduwe]@gmail.com}
}

\maketitle

\begin{abstract}
The Aho-Corasick algorithm is a multiple patterns searching algorithm running sequentially in various applications like network intrusion detection and bioinformatics for finding several input strings within a given large input string. The parallel version of the Aho-Corasick algorithm is called as Parallel Failure-less Aho-Corasick algorithm because it doesn’t need failure links like in the original Aho-Corasick algorithm. In this research, we implemented an application specific parallel failureless Aho-Corasick algorithm on the general purpose graphic processing unit by applying several cache optimization techniques for matching DNA sequences. Our parallel Aho-Corasick algorithm shows better performance than the available parallel Aho-Corasick algorithm library due to its simplicity and optimized cache memory usage of graphic processing units for matching DNA sequences.

keywords- Aho-Corasick, GPGPU, DNA sequence matching
\end{abstract}


%
\IEEEpeerreviewmaketitle

\section{Introduction}
The Aho-Corasick algorithm is used for finding multiple patterns (strings) within a given large input string. This was invented by Alfred V. Aho and Margaret J. Corasick \cite{Aho_original}. Therefore, this algorithm is used widely in many applications like plagiarism detection, intrusion detection, digital forensic, text mining and bioinformatics. The Aho-Corasick algorithm is executed one time for finding all occurrences of known patterns within the given input string. However, the main drawback of the original Aho-Corasick algorithm, which is based on sequential processing logic is the performance. The solution for the issue about the performance was introduced by Lin et al. \cite{pfac_1} \cite{pfac_2} by introducing a parallel version of the Aho-Corasick algorithm called Parallel Failure-less Aho-Corasick (PFAC) algorithm which runs on GPGPUs.

The original PFAC was introduced as the general purpose library for any kind of applications, which are based on finding multiple patterns within the given large input. However, our implementation is designed especially for DNA sequence matching and compared with the usage of the original PFAC for DNA sequence matching. Within our PFAC implementation, we concentrated on cache optimizations of GPGPU and application specific modifications for achieving better performance than original PFAC for DNA sequence matching.

\section{Related Work}
The Aho-Corasick algorithm was invented by Aho, Alfred V. and Corasick, Margaret J. on 1975 as an article ``Efficient String Matching: An Aid to Bibliographic Search" \cite{Aho_original}. They have introduced a new mechanism for searching multiple patterns within a given large input text. It shows good performance for multiple pattern searching among other traditional multiple pattern matching algorithm.

Dimopoulos et al. \cite{aho_corasick_memory_efficient} have discussed a modification to the serial Aho-Corasick algorithm for gaining memory efficiency to detect intrusions. They implemented a split Aho-Corasick algorithm with domain specific characteristics of intrusion detection for minimizing the memory usage of the finite state machine of the algorithm. As the domain specific characters, they observed that most patterns are subset of 256 characters, out of these 256 characters some are used almost in every states while other characters are used infrequently, and  split finite state machines have smaller memory sizes according to the domain than the large finite state machine. The modified algorithm has been run on a Field-Programmable Gate Array (FPGA) chip as an improved version of the Aho-Corasick algorithm with the point of view of the memory usage.

The speed-up of Aho-Corasick pattern matching machines by rearranging states has been done by Nishimura et al. \cite{aho_corasick_speed_up}. They have rearranged the states of the finite state machines for improving the memory access via cache memory. After constructing the basic goto graph, all the states of the graph have been rearranged according to the breadth-first order. Then all the memory accesses were more cache friendly and the performance gain is 55\%.

The PFAC was introduced by Lin at el.\cite{pfac_1}. PFAC has been run on a graphic processing units without failure links. They have run failure links less Aho-Corasick algorithm using thousand of threads of the GPGPU. Then it shows big performance improvement over the original serial Aho-Corasick algorithm. They published the second paper \cite{pfac_2} of improving the PFAC by introducing a hash function method.
 \IEEEpubidadjcol 

Kouzinopoulos et al. have done a survey about the performance of some selected string matching algorithm on GPGPU \cite{string_matching_on_GPGPU}. They have selected the Naive, Knuth-Morris-Pratt, Boyer-Moore and Quick-search on-line exact string matching algorithms as the test cases. However, the graphic card used for testing these algorithm was basic general graphic card called NVIDIA GTX 280. Therefore, they failed to show actual performance gain of these algorithm on the high end GPGPU like tesla series.

\section{Aho-Corasick algorithm}

As we mentioned, Alfred V. Aho and Margaret J. Corasick \cite{Aho_original} introduced the Aho-Corasick algorithm. They have discussed basic two steps for using this algorithm like (1). constructing a finite state pattern matching machine from the keywords and (2.) matching the input using the constructed finite state machine.

\begin{figure}[!t]
  \centering
  \includegraphics[width=3.2in]{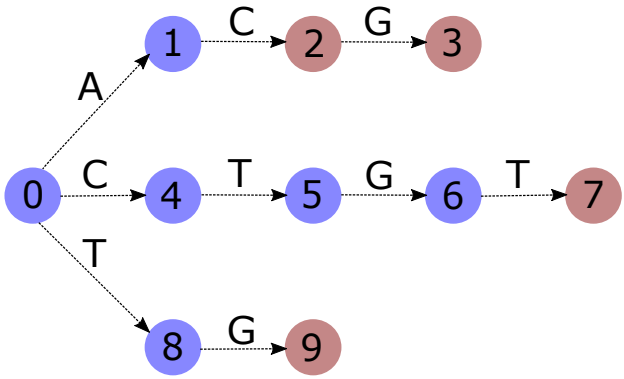}
  \caption{The goto graph for the patterns: AC, ACG, CTGT and TG}
  \label{fig:1}
\end{figure}

Let us look at an example; assume that patterns \textbf{AC, ACG, CTGT} and \textbf{TG} should be searched within the given text. As the first step, the finite state machine should be constructed using the given patterns. For constructing a finite state machine, they have used three functions called \textbf{goto (g), failure (f)} and \textbf{output}.
First, goto graph should be constructed for finding the goto function. This goto graph of the above example input patterns is given in Fig. \ref{fig:1}.

\begin{figure}[!t]
  \centering
  \includegraphics[width=3.2in]{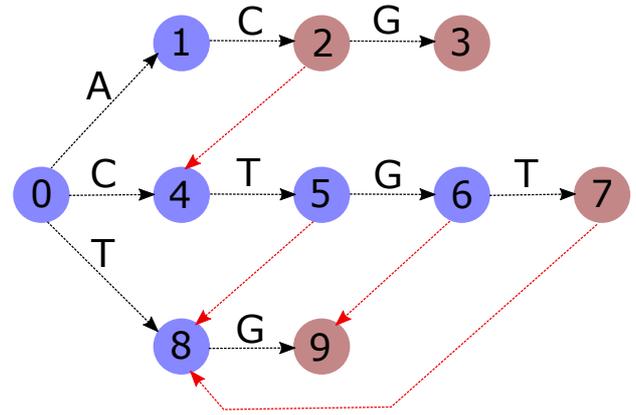}
  \caption{Transition diagram of the original Aho-corasick algorithm with failure links for the patterns AC, ACG, CTGT and TG}
  \label{fig:2}
\end{figure}

The pattern \textbf{AC} will be matched in state 2, \textbf{ACG} in state 3, \textbf{CTGT} in state 7 and \textbf{TG} in state 9 as shown in Fig. \ref{fig:1}. This graph represents the goto function.
Next step is constructing failure function from the goto function. This failure function should be calculated for all the states of the goto graph  as mentioned in the original Aho-Corasick article \cite{Aho_original}. The complete finite state machine with failure links of the above example is given in Fig. \ref{fig:2}. Some nodes are marked with red color to identify \textbf{output} nodes.

The complete finite state machine is now available as in Fig. \ref{fig:2}. It is easy to understand the processing phase of the Aho-Corasick algorithm by an example. Let's consider an input text with a text string S = s1,s2,s3,...,sn. The processing phase starts with the first state from the transition graph and the first character from the input text. First, it checks that the current input character of the input text has a transition from the current state of the finite state machine. Simply, it executes the \textbf{goto} function. If the \textbf{goto} function returns a new state, the current state of the input transition table is a new state while the current input character is the next character of the input.
Assume that the \textbf{goto} function failed, then the processing phase will check the \textbf{failure function (f) }for finding the next state for the current input character. If the finite state machine reached to the output nodes, it will return matched patterns of the current state of the finite state machine. All of the above steps will be continued until the process reaches to the final character of the input text.

\section{What is the PFAC Algorithm?}
Within this section, we are discussing Parallel Failure-less Aho-Corasick Algorithm (PFAC) implemented by Lin C.H. et al. \cite{pfac_1} \cite{pfac_2}. It works as same as the Aho-Corasick algorithm. However, PFAC runs on GPGPUs without failure links. PFAC doesn't use these failure links to backtrack the next starting point and sub patterns. It uses a parallel execution technique using parallel threads of GPGPUs. Then, it doesn't need any failure link for identifying the next start point because each character of the input has it's own thread for running a finite state machine. However, normally PFAC can detect only the longest patterns and it does not detect sub-patterns of the large patterns.

\begin{figure}[!t]
  \centering
  \includegraphics[width=3.3in]{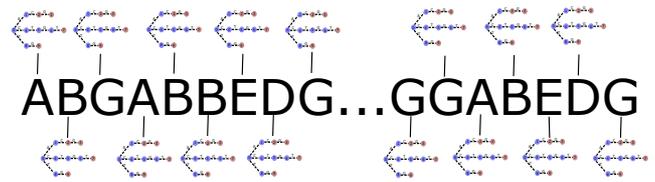}
  \caption{Parallel execution of the PFAC on the GPGPU}
  \label{fig:3}
\end{figure}

This failure-less machine runs parallel on a GPGPU. When the machine is run parallel on the GPGPU, each and every letter of the input text pass into the transition machine by it's own thread. As the result, every letter is searched in parallel individually. This mechanism can be identified by Fig. \ref{fig:3}.

PFAC algorithm has been used as the reference point of our research. We have developed our own application specific PFAC source code for matching DNA sequences because original PFAC was developed for general purposes. The development process of our application specific PFAC will be discussed in the next section.

\section{Designing and Implementation of the application specific PFAC for DNA sequence matching}
Within this section, we are discussing that how the new application specific PFAC algorithm was developed and features of that algorithm. Our own application specific PFAC algorithm  was developed with some differences from the original PFAC algorithm. However, our implementation has been limited into an application specific algorithm like bioinformatics DNA matching as a factor for gaining better performance than the general purpose original PFAC algorithm.

\subsection{Designing application specific PFAC algorithm for DNA sequence matching}
To develop an application specific PFAC algorithm, the following main steps were followed.

\begin{itemize}
  \item Developing a data structure for the input patterns' finite state machine.
  \item Read data from input text and run with the finite state machine.
  \item Managing memory types of the system for gaining better performance.
\end{itemize}

When the input pattern's finite state machine data structure was created, only four characters were considered because the bioinformatics algorithms of DNA matching are using mainly a specific number of letters like A,T,C and G. Therefore, 4$\times$N 2D array was created to store all the input patterns where N is the maximum number of states that can be handled by the program. A sample data structure (transition table) for storing sample input patterns can be seen at Table \ref{tbl:1}.

\begin{table}[!t]
\centering
\caption{A sample transition table for patterns AAATCG, TACGCC and AAATTG of our application specific PFAC implementation}
\label{tbl:1}
\begin{tabular}{lllll}
\hline
 & A & T & C & G \\
 \hline
0 & {\color[HTML]{FE0000} 1,0} & {\color[HTML]{FE0000} 6,0} & 0,0 & 0,0 \\
1 & {\color[HTML]{FE0000} 2,0} & 0,0 & 0,0 & 0,0 \\
2 & {\color[HTML]{FE0000} 3,0} & 0,0 & 0,0 & 0,0 \\
3 & 0,0 & {\color[HTML]{FE0000} 4,0} & 0,0 & 0,0 \\
4 & 0,0 & {\color[HTML]{FE0000} 11,0} & {\color[HTML]{FE0000} 5,0} & 0,0 \\
5 & 0,0 & 0,0 & 0,0 & {\color[HTML]{FE0000} 6,1} \\
6 & {\color[HTML]{FE0000} 7,0} & 0,0 & 0,0 & 0,0 \\
7 & 0,0 & 0,0 & {\color[HTML]{FE0000} 8,0} & 0,0 \\
8 & 0,0 & 0,0 & 0,0 & {\color[HTML]{FE0000} 9,0} \\
9 & 0,0 & 0,0 & {\color[HTML]{FE0000} 10,0} & 0,0 \\
10 & 0,0 & 0,0 & {\color[HTML]{FE0000} 11,2} & 0,0 \\
11 & 0,0 & 0,0 & 0,0 & {\color[HTML]{FE0000} 12,3} \\
12 & 0,0 & 0,0 & 0,0 & 0,0\\
\hline
\end{tabular}
\end{table}

Input patterns are loaded into this transition table one by one. Next state of the transition table is decided by current letter of the pattern and the previous state of the transition table. When the program find new state (new character) then it adds a new state to the transition table.

For an example consider the first pattern \textbf{AAATCG}  of the above example. First letter is \textbf{A}, then the program place next state number and matched pattern number within the Row=0 and Column=A, it is equal to (1, 0). Then, it takes the second character \textbf{A}. Same time, the pointer of the transition table moves to the new state. The loop finds the cell for letter \textbf{A}. It is equal to the cell (1, A). Next state and matched pattern number are placed within this cell. This mechanism is continued until it gets the final character of the current input pattern. Then, it will move to the next pattern.

Next step is, we have to read the input text into a buffer. After this step, the pattern matching machine should be run in parallel using a GPGPU with input text buffer same as the original PFAC execution as mentioned in early.

\subsection{Implementation of our application specific PFAC on GPGPU}

The main method of our application specific PFAC for loading input patterns into the transitions table can be seen in Fig.\ref{fig:code1}. It has three input arguments for parsing pattern file name, pointing next state of the int array and pointing the matched pattern ID array.

\begin{figure}[!t]
  \centering

\begin{framed}
\begin{minted}
[
fontsize=\fontsize{10}{12}
]
{C}

int
readPatternsFromFileToTransitionTable(
  char * patternFile,
  unsigned int * nextState,
  unsigned int * matchedPatternId){
  ...
  }
\end{minted}
\end{framed}
\caption{Method for loading patterns into the transition table}
\label{fig:code1}

\end{figure}

Next step of the program after loading input patterns into the transition table is reading data from an input text and running the kernel. Therefore, it is fair to consider that how to load input text data into the program. The code sample for reading data from an input text file is given in Fig. \ref{fig:code2}.

\begin{figure}[!t]
  \centering

\begin{framed}
\begin{minted}{C}
char * readTextFromFile(
  char * inputFile,
  int* inputText_size_p){
  ...
  }
\end{minted}
\end{framed}
\caption{Reading input text file into a buffer}
\label{fig:code2}
\end{figure}

According to the above method, the method readTextFromFile takes two arguments for input text file name and an integer pointer to store input file size. Finally, a char pointer which has all the characters of the input text file is passed as the return value.

Now, all the required data to run PFAC algorithm are ready within the memory of the host machine. Therefore, those data have to be passed into the device memory before launching a kernel in the device. The cudaMemcpy() function has been used for copying data from the host to device as well as the device to the host in the final stage.

Then, the kernel can be launched by passing the relevant parameters. Only one kernel is launched in our PFAC. It creates 256 threads per block. The number of blocks is defined according to the input text file size. If one dimension grid is considered, then the number of blocks per dimension may exceed the maximum number of blocks per dimension limit of the device for large input size. Therefore, a two dimension grid was created when the input has more characters. Deciding about dimensions of the grid is an automated process. As a requirement, the number of threads that is equal to the number of characters within the input text are created within the GPGPU for running our application specific PFAC algorithm. Finally within the kernel, character by character search is done using an individual thread using the transition table of the input patterns.

The kernel function takes mainly five parameters. The \emph{nextState} and \emph{matchedPatternId} are memory locations that were allocated previously to store the information about the transition table. Then, it takes main input called \emph{inputText} which has to be searched. The output is the integer array which has been used to store output data after finding a match. Our implementation of PFAC uses the 1D array instead of 2D arrays. Therefore, the pitch is taken as an input parameter to determine the rows of arrays. It means the pitch size is equal to the row size of the transition table. In our case, this value is equal to the four (4).

\subsection{Managing memory types to gain better performance}
The GPGPU is a device with several memory types. Global, shared, L1 cache, L2 cache, texture and constant memories are names to identify those various memory types. However, the usage of these memory types is different while some of the memory types are out of control from the user. Therefore, it is very important to handle these memory types according to the requirement.

Initially, all the data structures (input text data, transition table arrays) are transferred into the main global memory. Then those data structures are step by step migrated to special memory locations for testing the performance of each memory types. The first test was run using only global memory. Then two arrays of transition table called \textbf{nextState} and \textbf{matchedPatternId} are transferred into the texture memory location. Within another test, the input text also transferred into the shared memory. As the first step of testing, all the test cases run for different input patterns with large input text. As the second step, input data has been changed for measuring the performance for various input text.

\section{Experimental setup}

Our main workstation has Intel(R) Core(TM) i7-6700K 4.00GHz CPU with 32GB RAM. The main GPGPU was 6GB NVIDIA Tesla C2075. Table \ref{tbl:gpgpu_cache_architecture} is used to tabulate cache information about our GPGPU because cache optimizations are discussed under this research.

\begin{table}[!t]
	\centering
	\caption{Tesla c2075 GPGPU's cache architecture-6GB global memory}
	\label{tbl:gpgpu_cache_architecture}
	\begin{tabular}{llll}
		\hline
		& \begin{tabular}[c]{@{}l@{}}Cache \\ size\end{tabular} & \begin{tabular}[c]{@{}l@{}}Cache \\ line size\end{tabular} & Description \\
		\hline
		L1 cache & \begin{tabular}[c]{@{}l@{}}48KB/\\ 16KB\end{tabular} & 128bytes & \begin{tabular}[c]{@{}l@{}}can be disable\\ by using\\ -Xptxas-dlcm=cg\\ compile flag\end{tabular} \\
		\begin{tabular}[c]{@{}l@{}}Shared\\ memory\end{tabular} & \begin{tabular}[c]{@{}l@{}}16KB/\\ 48KB\end{tabular} & 128bytes & \begin{tabular}[c]{@{}l@{}}can be used \\ manually\end{tabular} \\
		L2 cache & 768KB & \begin{tabular}[c]{@{}l@{}}128bytes/\\ 32bytes\end{tabular} & Unified cache \\
		\hline
	\end{tabular}
\end{table}

Our PFAC implementation and original PFAC implementation were tested for various patterns and inputs. Information about these test cases is tabulated in Table \ref{tbl:2} and \ref{tbl:3}.

\begin{table}[!t]
\centering
\caption{Tested input patterns}
\label{tbl:2}
\begin{tabular}{lll}
\hline
\begin{tabular}[c]{@{}l@{}}Input \\ Pattern Set \\ Name\end{tabular} & \begin{tabular}[c]{@{}l@{}}Number \\ of \\ Patterns\end{tabular} & \begin{tabular}[c]{@{}l@{}}Size of \\ the File\end{tabular} \\
\hline
PatternSet 1 & 1000 & 101.1KB \\
PatternSet 2 & 2000 & 202.2KB \\
PatternSet 3 & 3000 & 3034KB \\
PatternSet 4 & 4000 & 404.5KB \\
PatternSet 5 & 5000 & 505.6KB \\
\hline
\end{tabular}
\end{table}

\begin{table}[!t]
\centering
\caption{Tested input data sets}
\label{tbl:3}
\begin{tabular}{lll}
\hline
Input Data Sets & \begin{tabular}[c]{@{}l@{}}Number of \\ Patterns\end{tabular} & \begin{tabular}[c]{@{}l@{}}Size of \\ the File\end{tabular} \\
\hline
Input DataSet 1 & 750000 & 76MB \\
Input DataSet 2 & 1500000 & 152.1MB \\
Input DataSet 3 & 2250000 & 228.1MB \\
Input DataSet 4 & 3000000 & 304.2MB \\
Input DataSet 5 & 3750000 & 380.2MB \\
\hline
\end{tabular}
\end{table}

\section{Results and discussion}

First, it is required to identify performance difference between memory types of the GPGPU for our application specific PFAC. Then, the cache configurations of the Fermi GPGPU was changed to identify the effects for the performance. Finally, our PFAC implementation and the original PFAC were tested for identifying the performance gap.

\begin{figure}[!t]
\begin{tikzpicture}
\begin{axis}
[
	x tick label style={
		/pgf/number format/1000 sep=},
	ylabel=Time (s),
	enlargelimits=0.05,
	legend style={at={(0.5,-0.1)},
	anchor=north,legend columns=-1},
	ybar interval=0.7,
    xticklabels={Pattern Set 1,Pattern Set 2,Pattern Set 3,Pattern Set 4, Pattern Set 5},
    x tick label style={rotate=20,anchor=east},
    legend pos=north west,
    legend columns=2,
    legend style={at={(0.5,-0.2)},anchor=north,font=\fontsize{5.5}{5}\selectfont},
    legend cell align=left,
    height=5cm,
    width=8cm,
    grid=both,
    minor ytick={0.225,0.25,...,0.475},
    label style={font=\tiny},
   tick label style={font=\tiny}
]

\addplot table [x=a,y=b,col sep=comma]{plot_data_pfac.csv};
\addplot table [x=a,y=c,col sep=comma]{plot_data_pfac.csv};
\addplot table [x=a,y=d,col sep=comma]{plot_data_pfac.csv};
\addplot table [x=a,y=e,col sep=comma]{plot_data_pfac.csv};
\addplot table [x=a,y=f,col sep=comma]{plot_data_pfac.csv};
\addplot table [x=a,y=g,col sep=comma]{plot_data_pfac.csv};
\legend{PFAC GM only - two arrays for pattern file
,GM+SM - two arrays for pattern file
,GM+SM+Texture - two arrays for pattern file
,GM - one array for pattern file
,GM+SM - one array for pattern file
,GM+SM+Texture - one array for pattern file
}
\end{axis}
\end{tikzpicture}
\caption{Time taken by our PFAC implementation for the various memory types of the GPGPU and various input patterns}
\label{graph1}
\end{figure}
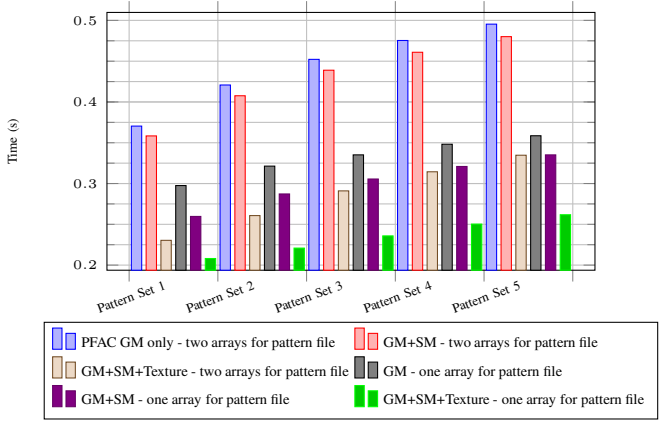

The first experiment of our application specific PFAC algorithm was done with only global memory. Then, shared memory and texture memory have been added into the algorithm gradually. The configurations of the GPGPU was unchanged. The test result of these tests is depicted in Fig. \ref{graph1}.


According to the result set in Fig. \ref{graph1}, it is clear that global memory to the GPGPU is the slowest memory as they mentioned in their official document. However, if the data structures are cache friendly then it is possible to gain considerable performance benefits. Within this test, array merge technique has been used to increase the spatial locality of the cache accesses.
Then, shared memory is applied into the accesses of input text pattern array. It improves the temporal locality of the cache accesses. Then, slight performance improvement could be gained from the global memory input text array compared the shared memory input text array. This performance gain is common for both cases which have two arrays for the transition table or one merge array for the transition table.

\begin{figure}[!t]
\begin{tikzpicture}
\begin{axis}
[
	x tick label style={
		/pgf/number format/1000 sep=},
	ylabel=Time (s),
	enlargelimits=0.05,
	legend style={at={(0.5,-0.1)},
	anchor=north,legend columns=-1},
	ybar interval=0.7,
    xticklabels={Pattern Set 1,Pattern Set 2,Pattern Set 3,Pattern Set 4, Pattern Set 5},
    x tick label style={rotate=20,anchor=east},
    legend pos=north west,
    legend columns=2,
    legend style={at={(0.5,-0.2)},anchor=north,font=\fontsize{5.5}{5}\selectfont},
    legend cell align=left,
    height=5cm,
    width=8cm,
    grid=both,
    minor ytick={0.225,0.25,...,0.475},
    label style={font=\tiny},
   tick label style={font=\tiny}
]

\addplot table [x=a,y=b,col sep=comma]{2_plot_2_data.csv};
\addplot table [x=a,y=c,col sep=comma]{2_plot_2_data.csv};
\addplot table [x=a,y=d,col sep=comma]{2_plot_2_data.csv};
\addplot table [x=a,y=e,col sep=comma]{2_plot_2_data.csv};
\addplot table [x=a,y=f,col sep=comma]{2_plot_2_data.csv};
\addplot table [x=a,y=g,col sep=comma]{2_plot_2_data.csv};
\legend{Default Config: - Two arrays transition table
,L1=48KB - Two arrays transition table
,L1=disabled - Two arrays transition table
,Default Config: - One array transition table
,L1=48KB - One array transition table
,L1=disabled - One array transition table
}
\end{axis}
\end{tikzpicture}
\caption{Time taken by our PFAC emplementation with shared memory for various cache configurations}
\label{graph02}
\end{figure}
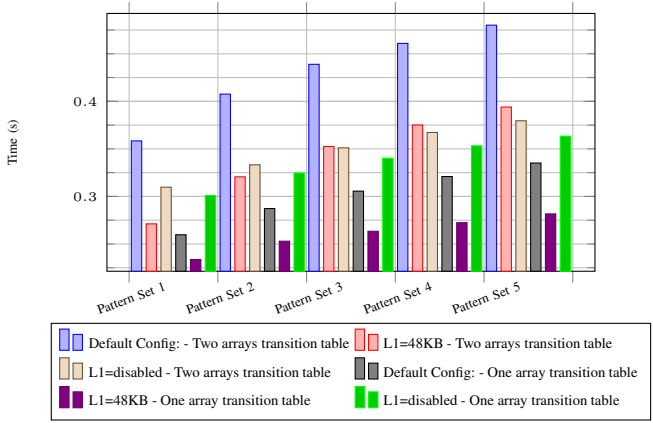

As the next step, texture memory has been applied to the memory accesses of the transition table. Then, transition table memory accesses were via the texture memory of the GPGPU. This increases the spatial locality and temporal locality of the memory accesses of the transition table because texture memory loads all the adjacent memory locations to the texture cache. The texture cache has a 2D spatial locality. Using this texture memory, it could be achieved better performance gain than the previous shared memory technique. However, the final test uses both shared memory and texture memory. At last, merged arrays with shared memory and texture memory shows the best performance among all other techniques.
Then, same tests like the above cases were done for the PFAC with shared memory because it shows the best performance among the tested methods. This modified PFAC accesses the input text via shared memory. In the previous basic PFAC, the input text is accessed via global memory only. The results of the PFAC with shared memory option are graphed in Fig. \ref{graph02}.


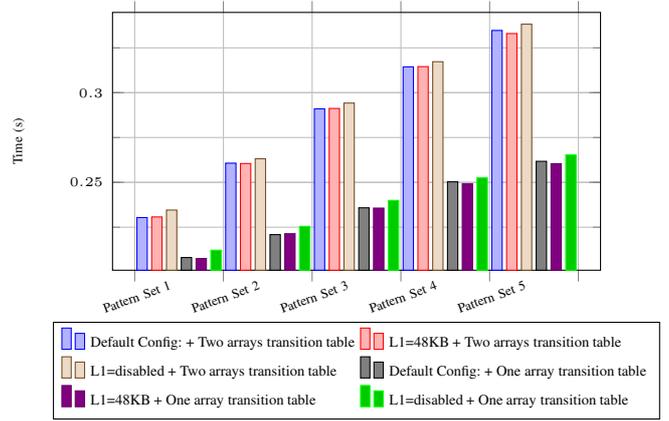
\begin{figure}[!t]
\begin{tikzpicture}
\begin{axis}
[
	x tick label style={
		/pgf/number format/1000 sep=},
	ylabel=Time (s),
	enlargelimits=0.05,
	legend style={at={(0.5,-0.1)},
	anchor=north,legend columns=-1},
	ybar interval=0.7,
    xticklabels={Pattern Set 1,Pattern Set 2,Pattern Set 3,Pattern Set 4, Pattern Set 5},
    x tick label style={rotate=20,anchor=east},
    legend pos=north west,
    legend columns=2,
    legend style={at={(0.5,-0.2)},anchor=north,font=\fontsize{5.5}{5}\selectfont},
    legend cell align=left,
    height=5cm,
    width=8cm,
    grid=both,
    minor ytick={0.225,0.25,...,0.475},
    label style={font=\tiny},
   tick label style={font=\tiny}
]

\addplot table [x=a,y=b,col sep=comma]{3_plot_3_data.csv};
\addplot table [x=a,y=c,col sep=comma]{3_plot_3_data.csv};
\addplot table [x=a,y=d,col sep=comma]{3_plot_3_data.csv};
\addplot table [x=a,y=e,col sep=comma]{3_plot_3_data.csv};
\addplot table [x=a,y=f,col sep=comma]{3_plot_3_data.csv};
\addplot table [x=a,y=g,col sep=comma]{3_plot_3_data.csv};
\legend{Default Config: + Two arrays transition table
,L1=48KB + Two arrays transition table
,L1=disabled + Two arrays transition table
,Default Config: + One array transition table
,L1=48KB + One array transition table
,L1=disabled + One array transition table
}
\end{axis}
\end{tikzpicture}
\caption{Time taken by our PFAC emplementation with shared and texture memory for various cache configurations}
\label{graph03}
\end{figure}

The shared memory of the GPGPU has been used in this experiment as the technique of optimizing memory access. As the results in Fig. \ref{graph02}, the PFAC with two array transition table shows better performance when it has the large L1 cache (48KB) than other options. The reason behind this story is reducing the cache replacements over the time. However, disabled L1 cache shows better performance if the input pattern set size is very large. This effect is occurring when the pattern size is large because L1 cache has to be replaced cache lines frequently time by time more compared with small input patterns. The number of cache lines within the L1 and the L2 cache is the main reason for this performance gain.
The PFAC with one array transition table and shared memory shows a pattern which is similar to the pattern of the previous experiment of the basic PFAC. Therefore, the reason behind this is the same as the above of giving different execution time for different cache options of basic PFAC with only global memory.
The texture memory has been used in next experiment for analysing the effect of cache memory options of the GPGPU within the PFAC algorithm. The modified PFAC uses the shared memory for the input text while the texture memory is used for the transition table of storing the input pattern set. The result set is shown in the Fig. \ref{graph03}.


According to Fig. \ref{graph03}. It is clear that there are no any clear differences between the 16KB L1 cache and 48KB L1 cache while the texture memory is used. The reason behind this effect is, the texture cache is completely different hardware cache located in each streaming multiprocessor of the GPGPU. The texture cache is designed especially for managing the 2D spatial locality. Therefore, it is well suited for the transition table of the modified PFAC algorithm. However, disabled L1 cache shows poor performance in the two array transition table and the one array transition table compared with enabled L1 cache because it is disabled the L1 cache completely. As the result, only shared memory is available for the usage within L1 cache location. The process cannot use the L1 cache for its own caching purpose. This introduces some performance losses.

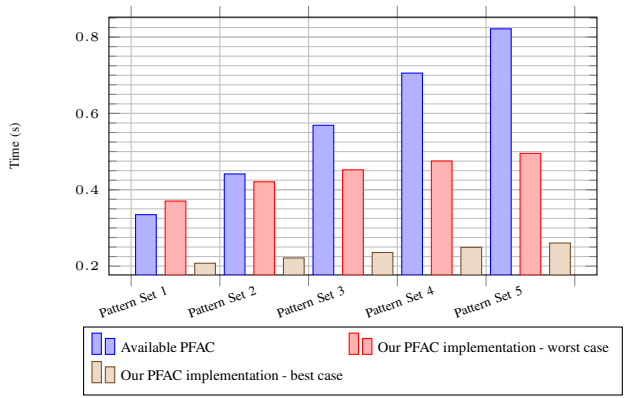
\begin{figure}[!t]
\begin{tikzpicture}
\begin{axis}
[
	x tick label style={
		/pgf/number format/1000 sep=},
	ylabel=Time (s),
	enlargelimits=0.05,
	legend style={at={(0.5,-0.1)},
	anchor=north,legend columns=-1},
	ybar interval=0.7,
    xticklabels={Pattern Set 1,Pattern Set 2,Pattern Set 3,Pattern Set 4, Pattern Set 5},
    x tick label style={rotate=20,anchor=east},
    legend pos=north west,
    legend columns=2,
    legend style={at={(0.5,-0.2)},anchor=north,font=\fontsize{5.5}{5}\selectfont},
    legend cell align=left,
    height=5cm,
    width=8cm,
    grid=both,
    minor ytick={0.225,0.25,...,0.9},
    label style={font=\tiny},
   tick label style={font=\tiny}
]

\addplot table [x=a,y=b,col sep=comma]{orginal_pfac_vs_our_pfac.csv};
\addplot table [x=a,y=c,col sep=comma]{orginal_pfac_vs_our_pfac.csv};
\addplot table [x=a,y=d,col sep=comma]{orginal_pfac_vs_our_pfac.csv};
\legend{Available PFAC
,Our PFAC implementation - worst case
,Our PFAC implementation - best case
}
\end{axis}
\end{tikzpicture}
\caption{Performance comparison between the original PFAC vs our PFAC implementation}
\label{graph04}
\end{figure}

A test has been done for checking the performance gap between the original PFAC and our PFAC implementation. The worst case and best case of our PFAC implementation were selected for comparing with available PFAC implementation. The results of this test are graphed in Fig. \ref{graph04}.


According to the result sets in Fig. \ref{graph04}, it is clear that our PFAC implementation has better performance than the available PFAC. The worst case of our PFAC implementation also is showing better performance than the available PFAC for large pattern sets. Reasons for this performance gap is different data structures which are more cache friendly and easy to handle compared with the data structures of the original PFAC because our PFAC implementation is application specific like DNA pattern matching.
If we compare the available PFAC and the best case of our PFAC implementations then a large gap between performances can be seen. For the largest data set, it is around 3X. The main reason behind this story is cache-friendly memory arrangements for running application specific PFAC compared with the original PFAC.

\begin{figure}[!t]
\begin{tikzpicture}
\begin{axis}
[
	x tick label style={
		/pgf/number format/1000 sep=},
	ylabel=Time (s),
	enlargelimits=0.05,
	legend style={at={(0.5,-0.1)},
	anchor=north,legend columns=-1},
	ybar interval=0.7,
    xticklabels={Data Set 1,Data Set 2,Data Set 3,Data Set 4, Data Set 5},
    x tick label style={rotate=20,anchor=east},
    legend pos=north west,
    legend columns=2,
    legend style={at={(0.5,-0.2)},anchor=north,font=\fontsize{5.5}{5}\selectfont},
    legend cell align=left,
    height=5cm,
    width=8cm,
    grid=both,
    minor ytick={0.25,0.5,...,3},
    label style={font=\tiny},
   tick label style={font=\tiny}
]

\addplot table [x=a,y=b,col sep=comma]{changing_data_pfac.csv};
\addplot table [x=a,y=c,col sep=comma]{changing_data_pfac.csv};
\addplot table [x=a,y=d,col sep=comma]{changing_data_pfac.csv};
\legend{Available PFAC
,Our PFAC implementation - worst case
,Our PFAC implementation - best case
}
\end{axis}
\end{tikzpicture}
\caption{Time taken by the available PFAC and our PFAC implementation for various input data}
\label{graph05}
\end{figure}
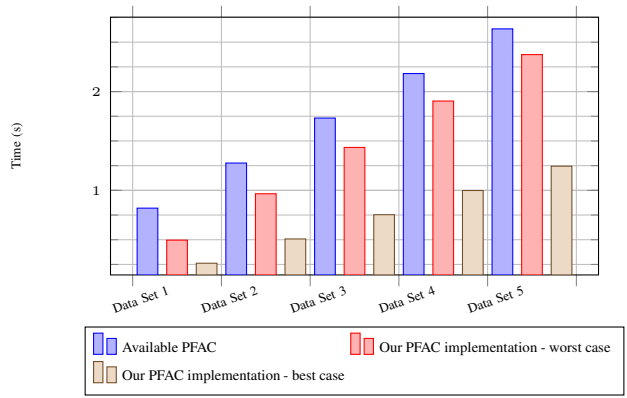

In all the previous tests, the variable factor was input patterns. As the next step, the input size of input has been changed for various sizes. However, data sets were changed only for identifying the effects of changing input data sets rather than changing the input pattern sets. Within this tests, all the cache parameters have not been changed one by one because the effect of cache options of the GPGPU has been tested in the above tests which were done for various input patterns. Therefore, only the worst case, best case and original available PFAC have been selected as the test cases for changing input data.
Then comparison between the available PFAC and our PFAC implementation was done with various input data files as mentioned in Table \ref{tbl:3}. The results of the experiment are graphed in Fig. \ref{graph05}.

Above tests were done for clarifying that our PFAC implementation has better performance while the input data are changed. According to the result sets in Fig. \ref{graph05}, our implementation has better performance for all the test input data in the worst case and the best case. The best case of our PFAC implementation has better performance for all the cases while it is 3X faster than the original PFAC for the largest data set which was used in our experiments. 

\section{Conclusion}

Our PFAC implementation with basic memory arrangement also shows better performance than the original PFAC. The original PFAC is better than our basic PFAC implementation when the input pattern size is small. The main reason behind this is, the original PFAC uses the texture memory of the GPGPU if the input pattern size is very small. They handle the texture memory automatically without enabling it all the time because they could not show better performance for large data with texture memory.

However, our PFAC implementation with all the possible cache optimization techniques shows the better performance than the original PFAC in all the time. Our best PFAC implementation uses the texture memory for handling the input data while shared memory for handling the input pattern. It is showing best performance among all the PFAC implementation because it has cache friendly data structures and access patterns.



%

\bibliographystyle{IEEEtran}
\bibliography{IEEEabrv,paper2}

\end{document}